\documentclass[10pt, conference, letterpaper]{IEEEtran}
\usepackage{cite}
\usepackage{amsmath,amssymb,amsfonts}
\usepackage[linesnumbered,ruled,vlined]{algorithm2e}
\usepackage{epsfig}
\usepackage{epstopdf}
\usepackage{multirow}
\usepackage{hhline}
\usepackage{verbatim}
\usepackage{algorithmicx}
\usepackage{lipsum}
\usepackage{algpseudocode}
\usepackage{graphicx,subfigure}
\usepackage{enumitem}
\usepackage{graphicx}
\usepackage{textcomp}
\usepackage{algcompatible}
\usepackage{relsize}
\usepackage{multicol}
\usepackage{multirow}
\usepackage{xcolor}
\usepackage[c3]{optidef}
\usepackage{setspace}

\newtheorem{property}{Property}

\definecolor{cadmiumgreen}{rgb}{0.12, 0.8, 0.17}
\usepackage{times}
\usepackage{fancyhdr}
\usepackage[ruled,vlined]{algorithm2e}

\newcommand{\Ec}{{\cal{E}}}
\newcommand{\Gc}{{\cal{G}}}
\newcommand{\Hc}{{\cal{H}}}

\newcommand{\Nc}{{\cal{N}}}
\newcommand{\Lc}{{\cal{L}}}
\newcommand{\Sc}{{\cal{S}}}
\newcommand{\vt}{\widetilde{v}}
\newcommand{\vh}{\widehat{v}}
\newcommand{\Vc}{{\cal{V}}}
\newcommand{\Vct}{\widetilde{\cal{V}}}
\newcommand{\Ect}{\widetilde{\cal{E}}}
\newcommand{\Gct}{\widetilde{\cal{G}}}

\usepackage{dblfloatfix}

\begin{document}
\title{Resource-aware Deployment of Dynamic DNNs \\ over Multi-tiered Interconnected Systems \vspace*{-3mm}}%
\author{C. Singhal$^{1}$, Y.~Wu$^{2}$, F.~Malandrino$^{3,4}$, M.~Levorato$^{2}$, C.~F.~Chiasserini$^{5,4,6}$\\
1: Indian Institute of Technology Kharagpur, India -- 2: 
UC Irvine, USA -- 3: CNR-IEIIT, Italy\\ 4: CNIT, Italy -- 5: Politecnico di Torino, Italy  -- 6: Chalmers University of Technology, Sweden
\vspace*{-0.5 cm}
}
\maketitle

\begin{abstract}
The increasing pervasiveness of intelligent mobile applications requires to exploit the full range of resources offered by the  mobile-edge-cloud network for the execution of inference tasks. However, due to the heterogeneity of such  multi-tiered networks, it is essential to make the applications' demand amenable to the available resources while minimizing energy consumption. Modern dynamic deep neural networks (DNN) achieve this goal by designing multi-branched architectures where \emph{early exits} enable sample-based adaptation of the model depth. In this paper, we tackle the problem of allocating sections of DNNs with early exits to the nodes of the mobile-edge-cloud system. By envisioning a 3-stage graph-modeling approach, we represent the possible options for splitting the DNN and deploying the DNN blocks on the multi-tiered network, embedding both the system constraints and the application requirements in a convenient and efficient way. Our framework -- named Feasible Inference Graph (FIN) -- can identify the solution that minimizes the overall inference energy consumption while enabling distributed inference over the multi-tiered network with the target  quality and latency. Our results, obtained for DNNs with different levels of complexity, show that FIN matches the optimum and yields over 65\% energy savings relative to a state-of-the-art technique for cost minimization.
\end{abstract}


\section{Introduction}

Deep Neural Networks (DNN) are a pervasive paradigm that empowers a wide range of  applications with advanced data analysis and decision making capabilities. Examples include computer vision \cite{li2021survey}, speech recognition \cite{nassif2019speech}, natural language processing \cite{goldberg2022neural}, and mobile health care \cite{azimi2017hich}. However, the complexity and energy consumption~\cite{zhang_wowmom23} of modern DNNs clashes with the limited computing power and energy reservoir of mobile platforms and embedded devices.

\begin{figure}[tb]
    \centering
    \includegraphics[width=3.5in]{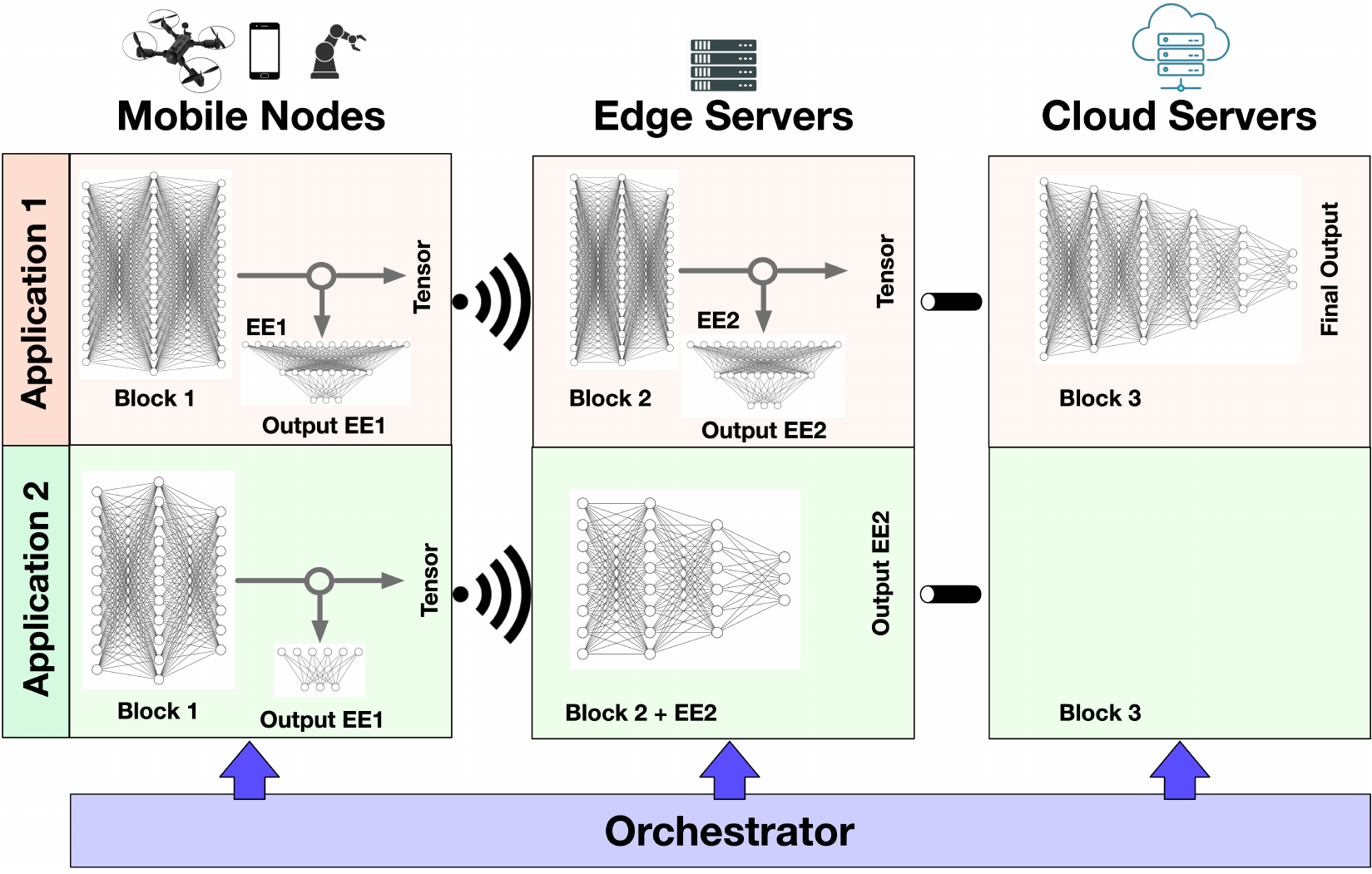}
        \vspace{-4mm}
    \caption{Our framework allocates blocks of layers of DNNs with early-exits (EEs) to mobile-edge-cloud systems so that energy expenditure is minimized. Multiple applications coexist, and an orchestrator controls the execution of DNN blocks and information flow  across them based on application requirements (accuracy and latency) and system constraints (bandwidth, computing capacity). For Application 1, the orchestrator allocates two blocks with early exits (EE1 and EE2) to a mobile node and an edge server, while for Application 2 the target performance is achieved by using the first two early exits.}
    \label{fig:intro}
    \vspace{-5mm}
\end{figure}

Two main technical strategies attempt to address this issue: model compression \cite{deng2020model} and edge computing \cite{luo2021resource}. In the former, the size and complexity of large DNN models are reduced using techniques such as pruning, quantization, and knowledge distillation. However, compression often results in a noticeable loss of inference performance, and in many mobile settings the execution of even compressed models necessitates a considerable amount of power. In edge computing, a compute-capable node positioned at the network edge takes over the execution of the DNN. Edge computing suffers from the need to transfer information-rich signals over a volatile wireless channel, often under tight latency constraints. Additionally, in real-world systems, edge servers likely need to support a multitude of applications and users, which may overload their computing and communications capabilities.

Recently, a variant of edge computing emerged where DNN models are divided into \emph{sections} that are distributed over  mobile-edge-cloud systems. This approach is often referred to as ``split computing'' or ``split DNN'' \cite{matsubara2022split}. 
{\em By controlling the splitting points defining the sections of the DNN, one can control the computing load allocated to the different devices/servers as well as the amount of data transmitted on the communication links connecting them.} In fact, each node is tasked with the execution of a portion of the DNN, and is thus in charge of a fraction of the overall operations. Moreover, instead of the input, the {\em sections transmit their output tensors, whose size is a function of the splitting point}. Relevant work in this direction includes \cite{kang2017neurosurgeon,yang2022edge,miao2020adaptive,lee2023wireless,fan2023joint}, as discussed in Sec.\,\ref{sec:relwork}.

In this paper, we tackle {\em the split DNN problem in the context of dynamic DNNs -- and, specifically, multi-branched DNNs equipped with early exits \cite{matsubara2022split,alexnet_ee,ee_sample_alexnet} -- and multi-tiered mobile-edge-cloud systems} (Fig.\,\ref{fig:intro}). The motivation behind these architectures is that, to achieve high performance in complex tasks, DNNs have to deal with the most challenging samples in a dataset. This results in models that are overparametrized for a large set of the input samples. Early exits are model-tails attached to some layers of the original model designed to produce an analogous output as that of the full DNN, with a smaller computing effort. By deciding which branches to execute sample-by-sample, early-exit models can dynamically adapt the number of operations needed to produce an output to the input ``complexity''. An  overview on early-exit DNNs and related challenges can be found in \cite{matsubara2022split,laskaridis2021adaptive}.

Our work focuses on the problem of allocating ``blocks'' of layers of DNNs with early exits to the nodes composing the overall mobile-edge-cloud system. Notably, unlike most of the existing studies on split DNNs, we consider a setting with multiple data sources, applications, and nodes at all the tiers of the infrastructure (see Fig.~\ref{fig:intro}). Also, it is worth remarking that the presence of early exits influences the flow of information throughout the blocks, and some executions may be terminated early, further complicating the allocation problem. The overall allocation problem we formulate minimizes the energy needed to complete inference under latency and accuracy application requirements, as well as  bandwidth and computing resource constraints. To resolve this challenging -- NP-hard -- problem, we adopt an approach based on graph optimization, where we manipulate an initial graph capturing the relationship between system nodes and DNN layers to create a specialized graph which we can be used to compute feasible low-cost paths (corresponding to allocation strategies).

In summary, the contributions of this work are as follows:
\begin{enumerate}
\item We formulate an allocation problem where the blocks of DNNs layers with early exits required by mobile  applications are allocated to the nodes of a mobile-edge-cloud system. The objective is to minimize energy consumption while satisfying application-specific inference requirements, under computing power and channel capacity constraints.
\item We devise a new solution framework -- named  Feasible Inference Graph (FIN) -- for the allocation problem. By manipulating the graph describing the DNN block allocation problem, we build a graph model amenable to   optimization, which  only contains paths 
representing feasible solutions and where the optimal allocation is the minimum-cost path. 
\item We explore the impact on the inference energy consumption of different allocation configurations for three DNN models with early exits (B-LeNet, B-AlexNet and B-ResNet, pre-trained on multiple datasets), over nodes with different computing and communication capabilities.
\item Our results show that models equipped with early exits can dramatically decrease the overall energy consumption when some of these exits are allocated to mobile or edge devices under system and application-level constraints, by reducing the involvement of larger-scale nodes in the completion of the inference. We also show how FIN performs close to the optimum computed by brute force.
Compared to the state-of-the-art approach in \cite{mcp1} for cost minimization, FIN yields  dramatic energy savings, exceeding 90\% of the computational energy and 80\% of the communication one. Furthermore, such gains are consistent under different branchy DNN architectures, from the small-scale B-LeNet to the much larger B-AlexNet.
\end{enumerate}

\begin{table}[tb]
\caption{Notation\vspace{-2mm}}
    \vspace{-3mm}
\small
\label{t:opt_param}
\begin{center}
\begin{tabular}{ |c|p{2.5in}| } 
 \hline
 \bf{Symbol} &  \bf{Definition} \\ \hline
 $s{\in}\Sc$ &  Data sources\\ \hline
 $n{\in}\Nc$ &  Multi-tiered network nodes\\ \hline
 $h{\in}\Hc$ &  Applications (or, equivalently,  DNN models)\\ \hline
 $\Gct{=}\{\Vct,\Ect\}$ &  Two-dimensional two-plane graph modeling the overall system\\ \hline
 $\Gc{=}\{\Vc,Ec\}$ &  Single-plane extended graph\\ \hline
$ \ell^h_i$ &  $i$-th block  of application $h$'s DNN \\\hline
  $\sigma^h$ &  Inference  rate for application $h$\\ \hline
$ \alpha^h$ &  Target inference quality of application $h$\\ \hline
$ \delta^h$ &  Target inference latency of application $h$\\ \hline
$\phi^h(\ell^h_i)$ &  Fraction of input samples that are output by $\ell_i^h$\\ \hline
 $T^h(v,v^{\prime})$,  &   Data transfer time, computing  time, and  energy  \\
 $C^h(v,v^{\prime})$, & consumption weights of the edge  $v{\rightarrow}v^\prime {\in} \Ec$, with \\ 
 $E^h(v,v^{\prime})$ & $v{=}(n,\ell^h_i)$ and $v^\prime{=}(n^\prime,\ell^h_j)$\\
\hline
$ \pi^h$ &  Path on $\Gc$ representing a configuration of application $h$\\ \hline
 $\gamma$ &  Resolution of the feasibility graph\\  \hline
 \end{tabular}
\end{center}
\vspace{-5mm}
\end{table}

\section{Energy-aware Inference through Dynamic NNs\label{sec:system}}

First, we introduce the  mobile-edge-cloud system, along  with the structure of the DNN models and applications. We also describe the corresponding two-plane graph capturing the possible allocation of DNN blocks (Plane 1) to the system nodes (Plane 2), along with the nodes computing and communication resources  (Sec.\,\ref{sub:system-model}). Then we translate the two-plane graph into a single-plane extended graph (Sec.\,\ref{sub:ext-graph}), used to   optimally configure  dynamic DNNs for energy-efficient inference tasks (Sec.\,\ref{sub:problem}).  We later develop a low complexity solution using this extended graph that gives performance close to optimum. 

\subsection{System model\label{sub:system-model}}
We consider a multi-tiered communication and computing infrastructure composed of mobile nodes, edge servers, and cloud servers. The objective of the overall system is to support a set of mobile applications whose central component is a DNN  performing the analysis of the information that the mobile nodes  acquire through co-located data sources. Examples include computer vision models for object detection and image classification, as well as speech recognition tasks. 
The edge of the infrastructure  hosts a Machine Learning (ML) orchestrator that possesses knowledge of the applications, as well as some essential information about the capabilities of the mobile nodes and the data they can acquire. 

Formally, the main elements defining the system are:
    
$\bullet$    A set $\Sc$ of data sources (e.g., sensors)  indexed with $s{\in}\{1,\ldots,S\}$;  
    each data source samples a physical phenomenon extracting information that serves as input to an inference task for application $h$ at rate $\sigma^h\geq 0$. 
    
     $\bullet$ A set $\Nc$ of computationally-capable network nodes, including (i) mobile nodes connected to data sources, (ii) edge servers, and (iii) cloud servers. 
    
     $\bullet$ A set $\Hc$ of applications $h{\in}\{1,\ldots,H\}$, each associated with a DNN model equipped with early-exits. We describe the overall architecture of the DNN associated with application $h$ as composed of a set of layer blocks $\mathcal{L}^h{=}\{\ell^h_1, \ldots, \ell^h_N\}$, where {\em each block corresponds to a portion of the network backbone and at most one early exit}. Although the input rate of the application is $\sigma^h$,  early exits may ``capture'' some input and terminate execution;  we thus define $\phi^h(\ell^h_i)$ as the fraction of input samples that are output after block $\ell_i^h$. 
     Each  application $h$  has specific  requirements defined as 
    the target inference quality (e.g., target accuracy value) $\alpha^h$, and maximum  inference latency $\delta^h$. In the following, we often refer to applications and the DNN representing their essential component interchangeably.

Given application $h$, the ML orchestrator determines which blocks of the DNN should be executed and assigns them  to network nodes in such a way that  the application inference requirements are fulfilled. Specifically, according to the split computing paradigm, the network nodes can cooperatively execute the overall DNN: a node assigned a DNN block or a set of blocks will execute all the corresponding layers and send the output coefficients of their cut layer (tensor) to the nodes hosting the subsequent block(s). Note that a node may be allocated zero, one, or multiple exits, which are all executed.

We represent the overall system by means of a directed \emph{two-dimensional load-resource, two-plane graph} model~\cite{GOUVEIA201922,GOUVEIA2017908,Multilayer1}, with  one plane capturing the network nodes and the other the blocks of the DNN layers.
We establish a relationship between these two planes to represent the possible mapping between the communication and compute resource demand of the applications in $\Hc$ onto the resources made available by the multi-tiered network nodes in $\Nc$. Also,  the two-dimensional load-resource model matches the communication and computing resources handled by the system, i.e., offered by the network nodes and required by the applications. 

\begin{figure*}
\centering
\includegraphics[width=1\textwidth]{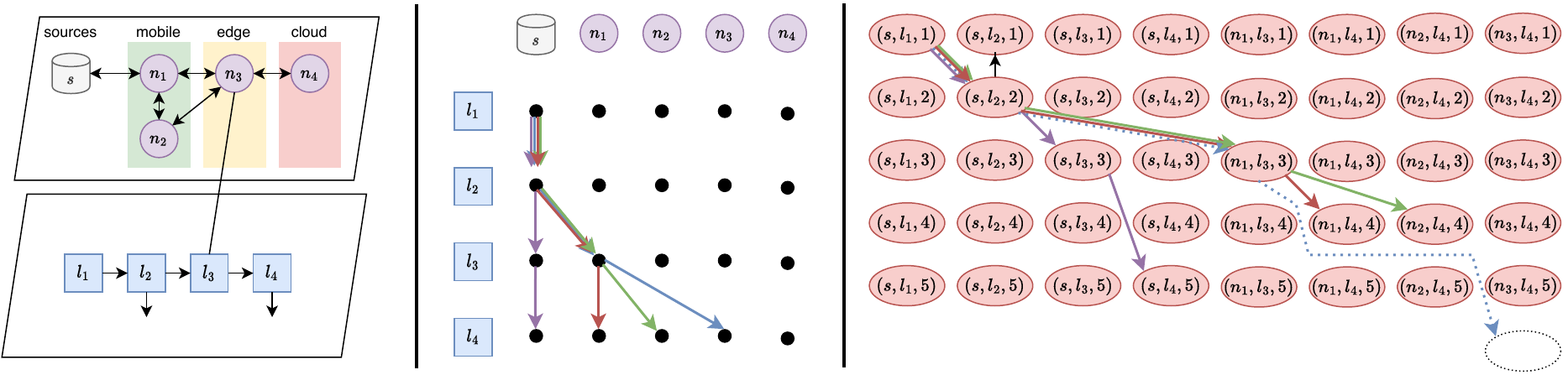}
    \vspace*{-4mm}
    \caption{
The graphs used in our solution strategy: two-dimensional, two-plane system model (left); single-plane extended graph  (center); feasible graph (right).
    \label{fig:fresco}
}
\vspace{-5mm}
\end{figure*}

More formally, we denote such graph, illustrated in the left panel of Fig.\,\ref{fig:fresco},
with $\Gct=\{\Vct,\Ect\}$ where $\Vct$ are the vertices and $\Ect$ are the edges. The two graph planes are as follows (the notation is summarized in Table\,\ref{t:opt_param}). \\
{\bf Plane 1:} The vertices $\Vct_1{\subset} \Vct$ of this plane correspond to the system nodes in $\Sc\cup\Nc$. A slicing setting is in place, where applications and nodes are assigned separate portions of computing resources and bandwidth. Thus, we associate with  the edges $\Ect_1{\subset}\Ect$ bidimensional weights $[b^h(n_1,n_2),c^h(n_1,n_2)]$, $n_1,n_2{\in}\Nc$, where $b^h(n_1,n_2)$ corresponds to the bandwidth of the communication link between nodes  $n_1$ and $n_2$ allocated to application $h$, and $c^h(n_1,n_2)$ is the computing power of  $n_1$ allocated to application $h$. The existence of the edge determines whether or not  two nodes can communicate, and the self loop $n{\rightarrow}n$ has infinite capacity, i.e., $b^h(n,n){=}\infty$. Further, the edge between a network node $n{\in}\Nc$ and a co-located data source $s{\in}\Sc$ has weight $[b^h(s,n),c^h(s,n)]=[\infty,0]$, i.e., we consider the bandwidth between the two $s$ and $n$ as unlimited and that the data source does not have any compute capability.
\\{\bf Plane 2:} The vertices $\Vct_2{\subset} \Vct$ of this plane correspond to DNN layers' blocks $\Lc{=}\{\Lc^h\}_h$. The edges $\Ect_2$ in this plane capture the connectivity structure of the DNNs, where edges exist only between consecutive blocks of the same application. The weights $[d^h(\ell^h_1,\ell^h_2),o^h(\ell^h_1,\ell^h_2)]$, $\ell^h_1,\ell^h_2{\in}\Lc^h$, represent the size (in bits) of the data output by  block $\ell^h_1$ ($d^h(\ell^h_1,\ell^h_2)$) and the number of operations needed to execute block $\ell^h_1$ ($o^h(\ell^h_1,\ell^h_2)$).\\
{\bf Inter-plane edges:} A set of edges $\Ect_{\ell\rightarrow n}$ unidirectionally connects Plane 2 to Plane 1, with the generic edge  representing that a  layers' block  of a DNN  is deployed at a network node.

\subsection{From the  two-plane to the  single-plane extended graph\label{sub:ext-graph}}
We now transform the two-dimensional load-resource, two-plane graph $\Gct$ into a directed  {\em single-plane} extended graph $\Gc=\left\{\Vc, \Ec\right\}$ (second panel of Fig.\,\ref{fig:fresco}). 
The vertices of $\Gc$ correspond to joint nodes and DNN blocks that are connected by inter-plane edges in $\Gct$, that is, in $\Gc$ the two planes are collapsed following the connecting edges. Furthermore, an edge in $\Gc$ exists only if the corresponding nodes and DNN blocks are connected in $\Gct$.
Consider the vertices $v{=}(n,\ell^h_i)$ and $v^{\prime}=(n^{\prime},\ell^{h}_j)$, where $n,n^{\prime}{\in}\Nc$ are network nodes and $\ell^h_i,\ell^h_j {\in}\Lc^h$ are  DNN blocks.  
Notably, vertices corresponding to data sources in $\Gc$ are not bound with any DNN block.

In $\Gc$, edges exist only between nodes embedding consecutive blocks of the same DNN (which may deployed also on the same node), and are associated with the set of weights $[T^h(v,v^{\prime}),C^h(v,v^{\prime}),E^h(v,v^{\prime})]$, where $T^h(v,v^{\prime})$ and $C^h(v,v^{\prime})$ are the data transfer time and the computing time, respectively, and $E^h(v,v^{\prime})$ is the per-inference energy consumption. We then define $T^h(v,v^{\prime})$ and $C^h(v,v^{\prime})$ as: 
\begin{equation}
T^h(v,v^{\prime}){=}\frac{d^h(\ell^h_i,\ell^h_j)}{b^h(n,n^{\prime})},~~
C^h(v,v^{\prime}){=}\frac{o^h(\ell^h_i,\ell^h_j)}{c^h(n,n^{\prime})}.
\end{equation}
We recall that $b^h(n,n^{\prime}){=}\infty$ if $n{=}n^{\prime}$, that is, the communication time is equal to $0$ if the vertices are associated with different (and necessarily contiguous) DNN blocks allocated on the same node. 
The weight $E^h(v,v^{\prime})$ instead compounds the computing, transmission, and receiving energy associated with edge $v{\rightarrow}v^{\prime}$, i.e.,
\begin{equation}\label{eq:energy}
E^h(v,v^{\prime})= (\xi_t + \xi_r) \frac{d^h(\ell^h_i,\ell^h_j)}{b^h(n,n^{\prime})} + \xi_c \frac{o^h(\ell^h_i,\ell^h_j)}{c^h(n,n^{\prime})},
\end{equation}
where $\xi_t$, $\xi_r$ and $\xi_c$ are the power consumption spent, respectively, by node $n$ to transmit, and by node $n^{\prime}$ to receive and compute, a data unit.

\subsection{Energy-efficient inference: Problem formulation\label{sub:problem}}

First, given application $h$, we denote with $\pi^h$ the generic configuration indicating (i) the sources feeding data to the application as well as (ii) {\em which}  DNN blocks with early exits should be deployed (hence used) for inference, and (iii) {\em where} (i.e., on which network nodes). We then define binary selection variables $p^h_{\pi}(v){\in}\{0,1\}$ based on the configuration $\pi^h$, where $v{\in}\Vc$. If $p^h_{\pi}(v){=}1$, then the vertex $v$ is selected by the configuration $\pi^h$, i.e., $v{\in}\pi^h$;  if $p^h_{\pi}(v){=}0$, $v$ is not selected, i.e.,  $v{\notin}\pi^h$. The configuration $\pi^h$ is controlled by the orchestrator. Note that the configuration needs to build a path from the sources to the DNN output for all the applications. Further, we remark that a node may be allocated multiple blocks, and even multiple exits. In the graph representation, this corresponds to a configuration that selects multiple vertices representing the same node and separate DNN blocks (like the red path in Fig.\,\ref{fig:fresco}(center)).

We adopt here a performance metric that is a function of the configuration and cannot be decoupled as a per-edge measure: the configuration inference quality. In fact, the configuration may suppress the execution of the blocks/exits after a certain index, for instance due to latency constraints. We thus define the inference quality of application $h$ given configuration $\pi^h$ as the quality associated with the whole sequence of DNN blocks in $\pi^h$, and denote it with $a(\pi^h)$.

The objective of the orchestrator is to minimize the overall energy consumption to support $\sigma^h$ (tasks per second) inferences per second \eqref{P2}, subject to latency, quality, network resource, and compute-resource constraints \eqref{C1_P2}--\eqref{C4_P2}. 
Recalling that  $v{=}(n,\ell^h_i)$ and $v^{\prime}{=}(n^{\prime},\ell^h_j)$, and $\phi^h(\ell^h_i)$ is the fraction of input samples output by $\ell^h_i$,  the resulting optimization problem (specified here for application $h$) is:  
\begingroup
\allowdisplaybreaks
\label{e:optimization}
 \begin{subequations}
\begin{align}
\min_{\pi^h} & \,\,\,  \sum_{v,v^{\prime} {\in} \Vc} \sigma^h \phi^h(\ell^h_i) E^{h}(v,v^{\prime}) p^h_{\pi}(v)p^h_{\pi}(v^{\prime})\label{P2}\\
 \textrm{s.t.} &\sum_{v,v^{\prime} {\in} \Vc}\left(T^h(v,v^{\prime}) + C^h(v,v^{\prime}) \right) p^h_{\pi}(v)p^h_{\pi}(v^{\prime}) \leq \delta^h \label{C1_P2}\\
 & a(\pi^h)\geq \alpha^{h} \label{C2_P2}\\
  & \sigma^h \phi^h(\ell^h_i)  o^h(\ell^h_i,\ell^h_j) \leq c^h(v,v^{\prime}), \,\,\,\forall v,v^{\prime}\in\pi^h\label{C3_P2}\\
 & \sigma^h \phi^h(\ell^h_i)  d^h(\ell^h_i,\ell^h_j) \leq b^h(v,v^{\prime}), \,\,\,\forall v,v^{\prime}\in\pi^h\,. 
 \label{C4_P2} 
\end{align}
\end{subequations}
\endgroup

\noindent
{\bf Problem complexity.} 
The above problem  is very complex to solve, owing to its combinatorial nature and to the overwhelming number of existing solutions. Specifically, we prove that the problem is NP-hard.
\begin{property}
The problem of optimizing \eqref{P2} subject to constraints \eqref{C1_P2}--\eqref{C4_P2} is NP-hard.
\end{property}
\begin{IEEEproof}
To prove NP-hardness, we perform a reduction from a known NP-hard problem to the one under study. 
In particular, we show that any instance of the Steiner tree problem (STP)~\cite{hwang1992steiner} can be transformed into a {\em simplified} instance of the problem introduced above. 
The STP is a generalization of the minimum spanning tree problem: given a weighted, undirected graph and a subset of nodes therein, the goal is to select the minimum-weight tree connecting all nodes in the subset.  Given an instance of the STP, we build an instance of the problem in \eqref{P2}--\eqref{C4_P2} by creating:
(i) a data source for all the vertices to connect except one, and imposing that all such sources must be used for inference;
(ii) one DNN layer, corresponding to the remaining vertex to connect;
(iii) physical nodes for all intermediate vertices in the STP instance;
(iv) only one of the physical nodes has enough capabilities to run the DNN layer. Also,
    the connectivity between nodes and data sources reproduces that of the STP instance, and 
   one component of the weights in our problem instance is set to match the weights in the STP instance while all others are set to zero.

Solving our problem to optimality also yields an optimal solution to the STP instance, hence, the two problems are equivalent. Since the reduction takes polynomial (linear) time (each edge and vertex of the STP instance is processed once) and  the STP problem is NP-hard~\cite{hwang1992steiner}, the thesis is proved.
\end{IEEEproof}
It is also worth remarking that the instance of our problem created in the proof above is very simple (only one DNN layer, only one non-zero weight for the edges, etc.). This suggests that, on top of being NP-hard, our problem is significantly {\em more} complex than an already NP-hard problem like the STP. In light of the problem complexity, we propose below an algorithmic solution, leveraging a graph representation that, efficiently and very conveniently, embeds all possible decisions, the application requirements, and the system constraints.

\section{The FIN Solution\label{sec:solution}}
Here, we  introduce our proposed heuristic, called  Feasible  Inference Graph (FIN). We  first describe how to build a {\em feasible} graph 
-- i.e., a graph summarizing all feasible solutions to the energy-aware inference problem -- 
starting from the extended one we used to formulate the  optimization problem.  
By construction, the feasible graph 
includes only  those decisions (i.e.,  which data sources and DNN blocks are used and where such blocks  should be deployed) that meet all  constraints on inference latency and quality as well as on data, computational, and network resources.  The second part of the section  then describes how the most energy-efficient  DNN  configuration can be found by identifying the minimum-cost path traversing the feasible graph, as the edge weights represent the energy consumption (computation and communication) incurred by the nodes.

Specifically, the feasible graph built by FIN summarizes 
all feasible solutions through two complementary strategies:
\begin{itemize}
    \item the {\em additive} constraint, namely, the inference latency, is guaranteed by the graph topology itself;
    \item the other constraints related to system capability and application requirements, e.g., inference quality and data  requirements as well as bandwidth limits, are guaranteed by pruning the edges and vertices that would violate them.
\end{itemize}

The vertices of the feasible graph are then the same as in the extended one, but each  vertex is replicated a number $\gamma$ of times. Let us denote the $g$-th replica of a vertex $v{=}(n,l^h_i)\mathord{\in} \Gc$ with $v_g$, $g{=}1,\dots,\gamma$, and define index~$g$ as {\em depth} of a vertex.
The vertices depth  is what allows us to track the additive constraints -- in our case, inference latency. Indeed, different replicas~$v_g$ of the same vertex~$v$ correspond to situations that are equivalent {\em except for the accumulated inference latency}:
intuitively, the deeper a vertex, the closer it is to violating  the additive constraint on the inference latency. Importantly, as mentioned earlier, all vertices correspond to solutions that do honor the latency limit; for vertices whose depth is exactly~$\gamma$ the constraint \eqref{C1_P2} is met with an equality sign.

Once the vertices are in place, we proceed to creating the edges connecting them. Specifically, we create an edge from vertex $v_{g_1}$ to vertex $v^\prime_{g_2}$, with $v^\prime{=}(n^\prime,l^h_j)$  and $g_2{>}g_1$, if:
\begin{itemize}
    \item it is possible to place layer~$\ell^h_i$ at node~$n$ and layer~$\ell^h_j$ at node~$n^\prime$;
    \item doing so incurs a combined processing time and network delay such that:
\end{itemize} 
    \begin{equation}
    \label{e:condition}
g_2-g_1 = \left \lceil  \gamma  \cdot \frac{T^h(v,v^{\prime})+C^h(v,v^{\prime})}
{\delta^h} \right \rceil.
    \end{equation}
In the numerator of the fraction above, the first term corresponds to the computing time, and the second to the data transfer delay. 

Let us then define the steepness of an edge as the difference between the depth of the target and the source vertices, and the steepness of a path as the sum of the steepness values of its edges. 
Intuitively, steeper edges (and steeper paths) correspond to solutions with a longer inference  latency; also, any path arriving to a vertex corresponding to an exit layer before depth~$\gamma$ represents, by construction, a solution conforming with the latency limit constraint.

Next, we prune all the vertices and edges that do not conform with the {\em local} constraints, i.e.,
\begin{itemize}
    \item edges that would not reach the target inference quality, i.e., they belong to a configuration $\pi^h$ s.t.\,$a(\pi^h){<}\alpha^h$, or
    \item edges that would exceed the available bandwidth or computational capability of the node represented by the vertex tail of the edge, i.e., 
   $\sigma^h \phi^h(\ell^h_i)  d^h(\ell^h_i,\ell^h_j) {>} b^h(v,v^{\prime})$ or
  $\sigma^h \phi^h(\ell^h_i)  o^h(\ell^h_i,\ell^h_j) {>} c^h(v,v^{\prime})$.
\end{itemize}
Surviving edges are assigned a weight corresponding to the energy consumption they incur, as defined in (\ref{eq:energy}). 
Thanks to the way the feasible graph is built and pruned, any path going from a data source vertex to any vertex corresponding to an exit layer corresponds to a feasible solution. To find the optimal one, it is thus sufficient to compute the minimum-cost path.

For instance, the right panel of Fig.~\ref{fig:fresco} depicts a feasible graph with $\gamma{=}5$. The red and green paths have steepness~$3$ and the purple one has steepness~$4$; they are all feasible and one of them will be selected as optimal solution. 
The blue path, instead,  {\em would} have a steepness of~$5$ and terminate at node~$(n_4,l_4,6)$; but this node does not exist in the feasible graph. The blue path as a whole is thus infeasible (hence, it is dotted in the figure), and is dropped from the graph.

As for parameter $\gamma$, this indicates the {\em resolution} with which the inference latency is represented by the feasible graph:   the smaller the $\gamma$ values, the fewer  the quantization levels (hence, the possible values of vertex depth and path steepness), and the larger the quantization error. As shown in Property~\ref{prop:optimalish}, such an error can be arbitrarily reduced by increasing $\gamma$, thus getting arbitrarily close to the optimum. 
\begin{property}
The competitive ratio of FIN (i.e., the ratio of the cost of FIN's solution to the cost of the optimal one) is~$1{+}\frac{1}{\gamma}$.
\label{prop:optimalish}
\end{property}
\begin{IEEEproof}
FIN consists of two steps: (a) building the feasible graph and (b) finding a minimum-cost path on it. Step (b) can be solved to optimality, e.g., through the Bellman-Ford algorithm. Step (a) can use the result of~\cite[Th.~4.3]{xue2007finding}, proving that creating~$\gamma$ replicas of the vertices in the feasible 
graph results in a cost increase of at most $\frac{1}{\gamma}$~times the optimum.
\end{IEEEproof}

On the negative side, a high value of $\gamma$ results in a higher complexity as the minimum-cost  traversal will search for more edges in the feasible graph. It is thus essential to set a value of $\gamma$ that effectively  trades off quantization  error with  complexity.  
Furthermore, given $\gamma$, we envision a $\lambda$-proximity approach to reduce the  complexity in searching for the vertices and edges to include in the output configuration. Specifically,  as the vertices with small  depth have a lower number of outgoing feasible edges, searching among vertices with depth close to $1$ may not help. We thus limit the search to a $\lambda$-proximity ($1{\leq} \lambda {\leq} \gamma$) index on the maximum depth vertices, i.e., with index $g{\in}[(\gamma{-}\lambda),\gamma]$, and $\lambda{=}\gamma$ corresponding to the exhaustive search among all nodes.

\begin{figure}[!htb]
    \centering
    \includegraphics[width=1\columnwidth]{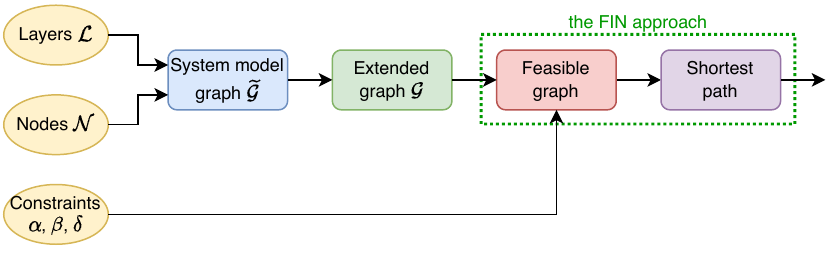}
        \vspace{-6mm}
    \caption{Solution strategy and steps within FIN.}
    \label{f:oslex}
    \vspace{-4mm}
\end{figure}

The steps involved in FIN are described in Alg.\,\ref{a:fin} and depicted in Fig.\,\ref{f:oslex}; the figure also shows how FIN is integrated with the construction of the two-plane system  graph and the single-plane extended graph.

\begin{algorithm}[!thb]
\small
\caption{FIN: Feasible graph and configuration selection}
\label{a:fin}
\begin{algorithmic}
\State \textbf{Input:} $\cal{G}$, $\alpha^h$, $\delta^h$,  $\gamma$, $\lambda$\\
 \SetKwFunction{FMain}{Main}
  \SetKwFunction{FSum}{\!\!}
  \SetKwFunction{FSub}{\!\!}
\noindent\SetKwProg{Fn}{Function I: Create\_feasible\_graph}{:}{}
  \Fn{\FSum{$\Gc$,$\gamma$}}{
  \begin{enumerate}     
     \item [1)] {\bf Create replica vertices}
\end{enumerate}
       ${}$\hspace{2.0em}\For{\text{each vertex} $v\in \Vc$}{
      \For{$g=0:1:\gamma$}{
      Create $v_g$ and include in feasible vertex set}}
      \begin{enumerate}     
     \item [2)] {\bf Create directed edges}
\end{enumerate}
      ${}$\hspace{0.0em} \For{\text{each edge} $(v,v^\prime)\in\Ec$}{
      \For{$g_1=0:1:\gamma$}{
      \For{$g_2=i:1:\gamma$}{
      \If{\eqref{e:condition} \mbox{holds true}}{
      Create edge $(v_{g_1},v^{\prime}_{g_2})$\;
      Include it in feasible edges set\;
      Compute edge weight $E_h(v_{g_1},v_{g_2}^{\prime})$ using \eqref{eq:energy}\;}
     }
     }
     }
    \KwRet feasible graph
  }  
\noindent\SetKwProg{Fn}{Function II: Configuration\_solution}{:}{}
  \Fn{\FSum{feasible graph, $\lambda, \gamma$ }}{
\begin{enumerate}
     \item Minimum-cost path traversal on feasible graph,\newline over $\lambda$-proximity vertices  \nonumber (i.e., $v_g$ with $g \in[(\gamma-\lambda),\gamma]$)
     \end{enumerate}
     ${}$\hspace{2.0em} Initialize $g_1=1$\\
     ${}$\hspace{2.0em} Include $v_{g_1}=(n,l^h_1)$ in $\pi^h$ \\
      ${}$\hspace{2.4em}\For{$i=2:|\Lc^h|$}{%
     \For{$g_2=(\gamma-\lambda):\gamma$}{%
     \If{edge $(v_{g_1},v^{\prime}_{g_2})\colon v_{g_1}=(n,l^h_{i-1})$ }
     {Include $v^{\prime}_{g_2}$ in $\pi^h$ if $\min\limits_{v^{\prime}}E^h(v,v^{\prime})$\;
     Set $g_1=g_2$;}
            }
            }
     
    \KwRet $\pi^h$ as FIN output configuration
  } 
\end{algorithmic}
\end{algorithm}

\section{Reference Scenario\label{sec:reference-scenario}}
{\bf Branched DNNs with early-exits.}
To evaluate the performance of our FIN solution, we consider six different applications corresponding  to three DNN models with early-exits, namely, B-LeNet, B-AlexNet, and B-ResNet, each trained with two popular datasets. We summarize such mapping  in Table\,\ref{t:application} and detail the three models  below. We also underline that we consider accuracy as measure of the inference quality. 

{\em B-AlexNet} is
a branchy version of AlexNet \cite{AlexNet1} DNN architecture with early-exits. 
It has 5 convolution, 1 max-pooling, 3 fully connected, and 3 early-exit blocks. It can be used for image  classification
\cite{AlexNet1},  video summarization 
\cite{alexnet_vid_sum}, and human activity classification 
\cite{alexnet_activity1,alexnet_activity2,alexnet_activity3,alexnet_ee}.  
The input to the B-AlexNet model is RGB format images, scaled to size 227$\times$227$\times$3. The feature map and complexity of the B-AlexNet architecture  is given in Table\,\ref{t:feature_map}; 
when trained with datasets from different sources and with different sample rate (i.e., samples/category in CIFAR10 and CIFAR100) \cite{ee_sample_alexnet}, it results in an accuracy level as given in Table\,\ref{t:accuracy_all}.



\begin{table}[!tbp]
\vspace{-2mm}
\caption{Details of the pre-trained DNN models with early exits  used by the six applications\vspace{-2mm}}
    \vspace{-2mm}
\label{t:application}
\begin{center}
\small
\begin{tabular}{|c|c|p{0.5in}|c|p{0.76in}|}
    \hline
    \!\!{{\bf Application}}\!\!& {\bf DNN} & {\bf Training dataset} & {\bf\#\,exits}\!\!& {\bf Exit\,output\,\,\,\,\,\,\, $\phi^h$\,[\%]}\\\hline
    $h_1$&\multirow{2}{*}{\!\!B-AlexNet\!\!}&CIFAR100&3&\multirow{2}{*}{\!\!\![65.6,\,25.2,\,9.2]}\\\cline{1-1}\cline{3-4}
    $h_2$&&CIFAR10&3&\\ \hline
    $h_3$&\multirow{2}{*}{B-ResNet}&CIFAR100&3&\multirow{2}{*}{\!\!\![41.5,\,13.8,\,44.7]}\\\cline{1-1}\cline{3-4}
    $h_4$&&CIFAR10&3&\\ \hline
    $h_5$&\multirow{2}{*}{B-LeNet}&MNIST&2&\multirow{2}{*}{[94.3,\,5.63]}\\\cline{1-1}\cline{3-4}
    $h_6$&&EMNIST&2&\\ \hline
 
\end{tabular}
\end{center}
\vspace{-5.5mm}
\end{table}

\begin{table}[!tbp]
\caption{Number of input features and complexity of the DNN model blocks\vspace{-2mm}}
    \vspace{-2mm}
\label{t:feature_map}
\begin{center}
\small
\begin{tabular}{|c|c|c|c|}    
    \hline
   {\bf Block}&\multicolumn{3}{|c|}{{\bf [Number of features, \,Complexity [MOPs]]}}\\\cline{2-4}
 &   B-AlexNet &  B-ResNet &  B-LeNet\\\hline
    1 & [290400,\,0.043] & [16384,\,0.004] & [4704,\,0.118] \\ \hline
    2&[186624,\,6.711] & [16384,\,0.021] & [1600,\,0.040]  \\ \hline
   3 & [64896,\,10.145]& [16384,\,0.021] & [120,\,0.048]   \\ \hline
    4 & [64896,\,13.523] &  [4096,\,0.083] &-  \\ \hline
    5 & [43264,\,29.045] & [4096,\,0.664]  &-\\ \hline
     Exit-1 & [64896,\,22.579] & [4096,\,0.748] & [120,\,0.05] \\ \hline
 Exit-2&[43264,9.056]& [4096,\,0.665]& [10,\,0.022]\\ \hline
    Exit-3 & [1000,\,0.039] & [10,\,0.001] &- \\ 
    \hline
\end{tabular}
\end{center}
\vspace{-5mm}
\end{table}

\begin{table}[!tbp]
\caption{Inference accuracy of the pre-trained DNN models with early exits used by the applications\vspace{-2mm}}
    \vspace{-2mm}
\label{t:accuracy_all}
\begin{center}
\small
\begin{tabular}{|c|c|c|c|c|c|c|}
    \hline
    {\bf Exit} & \multicolumn{6}{|c|}{\bf Accuracy [\%]}\\\cline{2-7}
    {\bf block}& $h_1$ &$h_2$&$h_3$&$h_4$&$h_5$&$h_6$\\ \hline
     Exit-1& 39.56& 56.37 & 29.97& 38.97 &91.18 &93.54\\ \hline
     Exit-2 & 54.22 &78.04& 39.93&  51.93 &96.70&99.20\\ \hline
    Exit-3 & 60.32&85.95&  72.21& 93.91 & -&-  \\ \hline
 
\end{tabular}
\end{center}
\vspace{-5mm}
\end{table}


{\em  B-ResNet} is   
a branchy version of ResNet110 \cite{resnet} DNN  with early-exits used for image recognition and classification tasks. ResNet110 consists of 3 stages, with each stage comprising a series of residual blocks. The first stage has 18 residual blocks with 16 filters in each block, while the second and third stages have 36 residual blocks, each with 32 and 64 filters in each block (resp.). The residual blocks  consist of 2 or 3 convolutional layers, each followed by batch normalization and ReLU activation. 
The input  is size 32$\times$32$\times$3, while the feature map and complexity of B-ResNet architecture layers are given in Table\,\ref{t:feature_map}. The inference accuracy of pretrained B-ResNet using CIFAR10 and CIFAR100 is given in Table\,\ref{t:accuracy_all}.

{\em B-LeNet} is a branchy version of LeNet-5 CNN architecture \cite{LeNet} that can recognize handwritten digits using datasets like  MNIST \cite{mnist} and EMNIST \cite{emnist}. 
B-Lenet consists of 8 layers (2 convolutional,  2 subsampling, 3 fully-connected,  1 early-exit). Its feature map and complexity  are given in Table\,\ref{t:feature_map}, while the inference accuracy of pretrained B-Lenet using MNIST \cite{mnist} and EMNIST \cite{emnist} is given in Table\,\ref{t:accuracy_all}. 

\begin{figure*}[!bh]
\vspace*{-4mm}
    \centering     
         \includegraphics[width=0.32\textwidth]{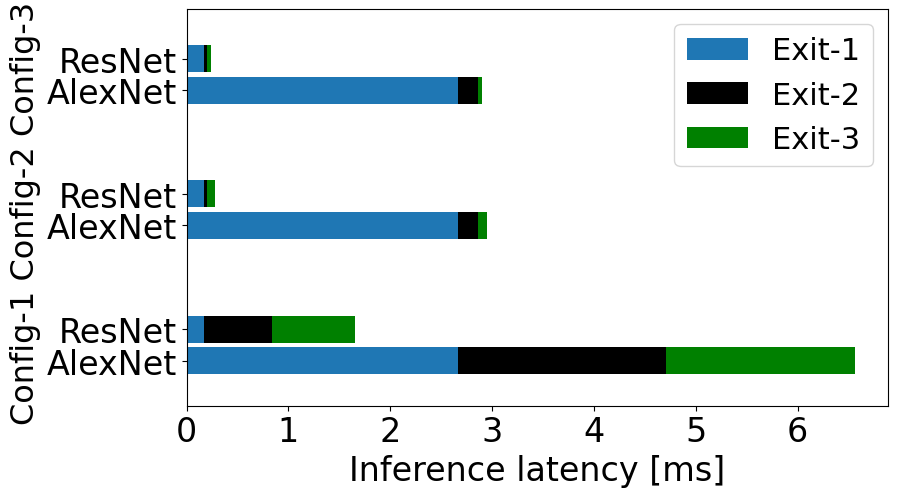} 
\hspace{1mm}
\includegraphics[width=0.32\textwidth]{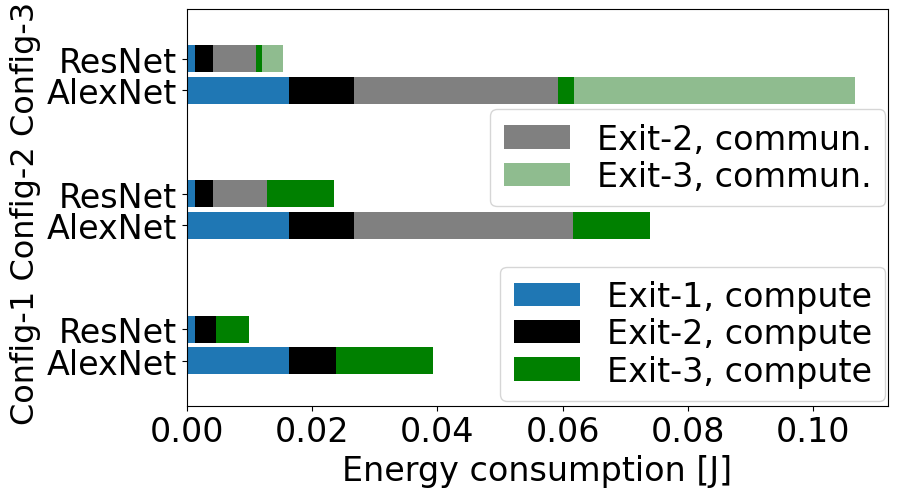}  
\hspace{1mm}
\includegraphics[width=0.32\textwidth]{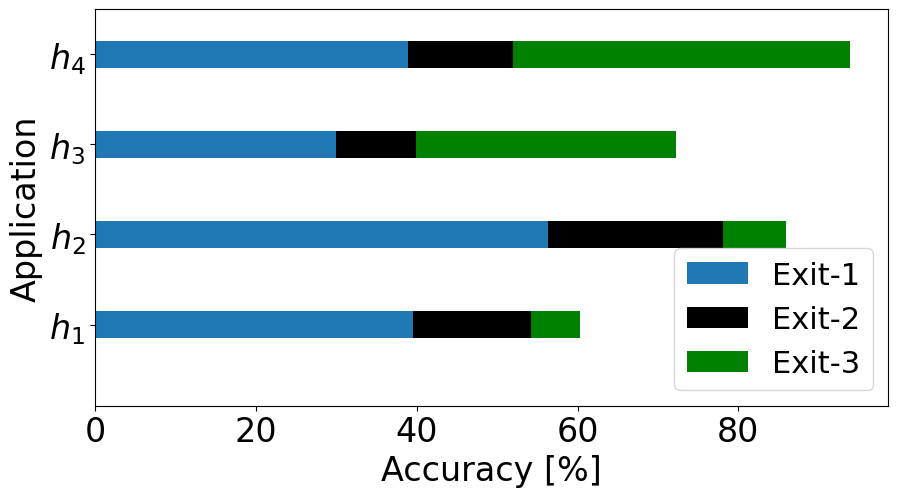}   
    \vspace*{-3mm}
    \caption{Impact of the configurations listed in Table\,\ref{t:config}: inference latency (left) and energy consumption (center) for the B-AlexNet and B-ResNet; inference accuracy for B-AlexNet-based $h_1$, $h_2$ and B-ResNet-based  $h_3$, $h_4$ (right).}
\label{f:res_config}
\end{figure*}

\begin{table}[!tb]
\caption{Communication (downlink (DL)/uplink (UL)) capacity and energy consumption of the system  nodes\vspace{-2mm}}
\vspace{-2mm}
\label{t:commun}
\begin{center}
\small
\begin{tabular}{|c|c|c|p{1.7cm}|p{2cm}|}
    \hline
    \multirow{2}{*}{\bf Node}&\multicolumn{2}{|c|}{\bf Power\,[W]}&\!\!{\bf Traffic\,DL/UL}&\!\! {\bf Energy\,DL/UL}\!\!\\ \cline{2-3}
    & Idle &Max&{\bf [Gbps]}\!\!&\!\!{\bf [nJ/bit]}\!\!  \\ \hline
    Mobile  & 3.1 & 3.7 & 0.1 & 30 \\ \hline
    Edge & 4,096 & 4,550 & 560 & 37\\ \hline
    Cloud  & 11,070 & 12,300 & 4,480 & 12.6 \\\hline
\end{tabular}
\end{center}
\vspace{-5mm}
\end{table}

{\bf Network system.}
We use three types of  nodes, i.e., mobile, edge, and cloud nodes \cite{compute_par}, respectively, associated with the following values of computational capability in trillions operations per second (TOPS) and power consumption in watt (W): [11\,TOPS,\,6\,W]; [153.4\,TOPS,\,140\,W], [312\,TOPS,\,400\,W]. 
The capability of the communication interface of such nodes are given in Table\,\ref{t:commun}\cite{commun_par1,commun_par2,commun_par3}. 
We recall that such  parameters define the  bandwidth and computing resource  weights of the edges between the vertices in Plane 1 of graph $\widetilde{\mathcal{G}}$, and the corresponding power consumption. Further, in our experiments, we initially consider one node for each network tier, and then we  let the number of mobile nodes increase with the number of users.

\section{Performance Evaluation\label{sec:results}}

In this section, we first show the impact of different DNN configurations on the inference latency, quality, and energy cost. Then, we introduce the alternatives against which we compare FIN, and present the performance obtained for the applications and network system described earlier. 

\subsection{Impact of the DNN configurations} 
We start by investigating the trade-off  among energy consumption, inference accuracy, and inference latency while dynamically orchestrating a DNN model on the multi-tiered    network for various user applications. 
To this end, we consider three example configurations for B-AlexNet and B-ResNet, in which the DNN blocks are deployed on the mobile, edge, and cloud nodes as indicated in Table\,\ref{t:config}. 

Fig.\,\ref{f:res_config} shows that, as expected, both Config-2 (involving mobile and edge) and Config-3 (involving all tiers) reduce the inference latency overall compared to Config-1 (involving the mobile only).  However, it is interesting to observe that the benefit of involving the cloud in the inference task (i.e., Config-3) is negligible when compared to Config-2 where only the edge is added in support to the mobile node. 

Looking instead at the energy consumption and the accuracy performance together, while Config-2 and Config-3 make the computing burden at the mobile nodes lighter, they may imply a higher overall cost due to the  communication energy expenditure. Such an expenditure appears   whenever exit-2 and exit-3 are enabled (i.e., a higher accuracy is required).  
The increase in communication energy is especially noticeable for B-AlexNet, which has indeed the largest size,  
resulting in a significant surge (87.6\%) in the overall cost when transitioning from Config-1 to Config-2, and a further increase of 28\% when shifting from Config-2 to Config-3.
 Instead, whenever a lower accuracy is acceptable, we can exploit the splits corresponding to exit-1 for all DNNs, and, so doing, reduce both inference latency (from 6.56\,ms to 2.67\,ms), and energy consumption (from 39.4\,mJ with all three exits active to just 16.4\,mJ when only exit-1 is activated in Config-1).

In summary, whenever the mobile nodes are used, even for a subset of the DNN blocks, the inference latency grows to such an extent that the reduction in computing time  brought by cloud nodes is negligible. Furthermore, whenever we aim at the maximum accuracy, using cloud nodes may lead to very high communication energy costs. Remarkably, however, there exist configurations involving nodes from the different network tiers that can reduce the total as well as the mobile node energy expenditure. It follows that it is possible to identify DNN allocation strategies that improve the sustainability of inference tasks when the application requirements warrant it.

\begin{table}[!tb]
\caption{Example test deployment configurations of B-AlexNet and B-ResNet\vspace{-2mm}}
    \vspace{-4mm}
\label{t:config}
\begin{center}
\small
\renewcommand{\arraystretch}{1.4}
\begin{tabular}{|cc|c|c|c|c|}
    \hline
     \multicolumn{3}{|c|}{\bf Configuration}&{\bf Mobile} &{\bf Edge} & {\bf Cloud} \\  \hline
   \parbox[t]{0.01mm}{\multirow{3}{*}{\rotatebox[origin=c]{90}{B-AlexNet}}} &\parbox[t]{1mm}{\multirow{3}{*}{\rotatebox[origin=c]{90}{B-ResNet}}}& Config-1 & All blocks&- &-\\ \cline{3-6}
   & & Config-2 & \!\!$\ell^h_{1}, $exit-1,$ \ell^h_{2},$ \!\!&\!\!$\ell^h_{3}, $exit-2,$ \ell^h_{4}, \ell^h_{5}, $exit-3\!\!&-\\\cline{3-6}
    & & Config-3 & \!\!$\ell^h_{1}, $exit-1,$ \ell^h_{2},$ \!\!&$\ell^h_{3}, $exit-2,$ \ell^h_{4}$ & $\ell^h_{5}, $exit-3\\\hline
\end{tabular}
\renewcommand{\arraystretch}{1}
\end{center}
\vspace{-5mm}
\end{table}

\subsection{FIN performance\label{sec:results}} 

\noindent
{\bf Benchmarks.}  We compare FIN against:

\textbullet~\emph{Multi-constrained path selection (MCP)}, a solution to our problem based on the multi-constrained path selection  in \cite{mcp1}. 
We select \cite{mcp1} because no scheme exists that specifically tackles the problem at hand. 
\cite{mcp1} finds a path between  source and destination nodes in a graph such that it satisfies the multiple end-to-end constraints on the additive edge weights.  
MCP applies such an approach to our extended graph $\Gc$ to find a feasible solution for the optimization problem in Sec.\,\ref{sub:problem}. To this end, we assign to  edge  $v {\rightarrow} v^{\prime} {\in} \Ec$ the   auxiliary weight:    
$\Omega(v,v^{\prime}){=}(\frac{T^h(v,v^{\prime}){+} C^h(v,v^{\prime})}{\delta^h}{+}\frac{a(v^{\prime})}{\alpha^h})$, where $a(v^{\prime})$ is the accuracy 
 associated with the whole sequence of DNN blocks till $v^{\prime}$. 
Then, the minimum-cost path is selected using the auxiliary edge weights in the extended graph. 

\textbullet~\emph{Optimum (Opt),} 
 obtained through  exhaustive search.  

\begin{figure}
    \centering
      \subfigure[$\alpha^h=$80$ \%$ \label{f:tot_energy_balexnetb}]{
      \includegraphics[width=0.475\columnwidth]{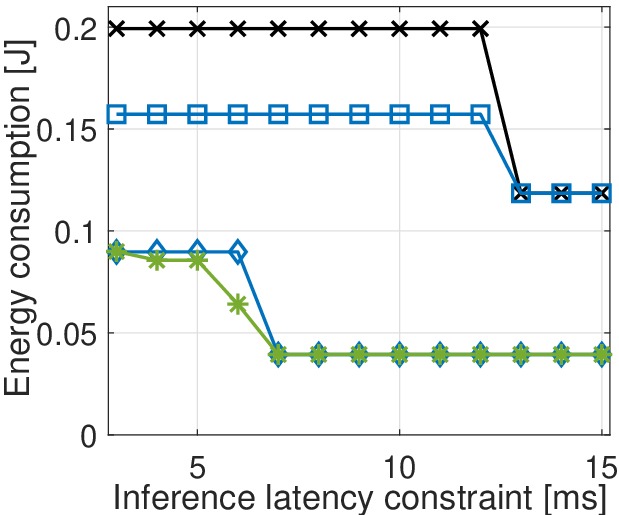}  
    }\hspace{-2mm}
         \subfigure[$\delta^h=$5 ms\label{f:tot_energy_balexneta}]{
         \includegraphics[width=0.475\columnwidth]{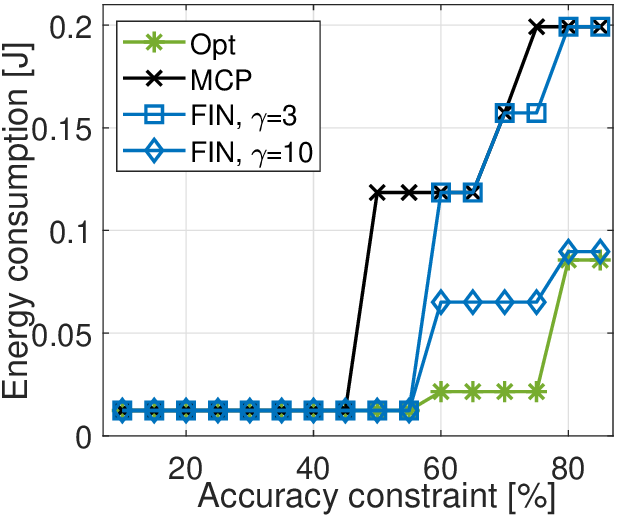}
         }
             \vspace{-3mm}
    \caption{Total energy consumption of the B-AlexNet configurations obtained through Opt, MCP, and FIN  ($\gamma{=}3,10$),  as the target inference  latency and accuracy vary.}
    \label{f:tot_energy_balexnet}
    \vspace{-5mm}
\end{figure}

\begin{figure}[!b]
    \centering
         \subfigure[$\alpha^h{=}80$\%, commun. energy\label{f:all_config_constraintb}]{
                       \includegraphics[width=0.455\columnwidth]{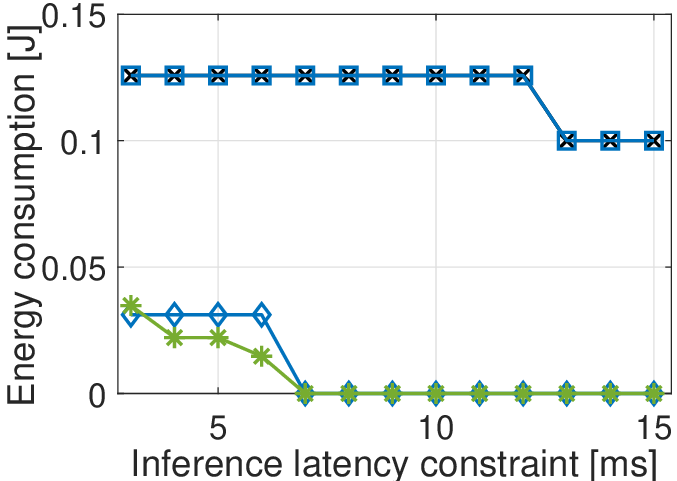}
                       }     \hspace{-2mm}            
     \subfigure[$\delta^h{=}$5\,ms, commun. energy\label{f:all_config_constrainta}]{
  \includegraphics[width=0.455\columnwidth]{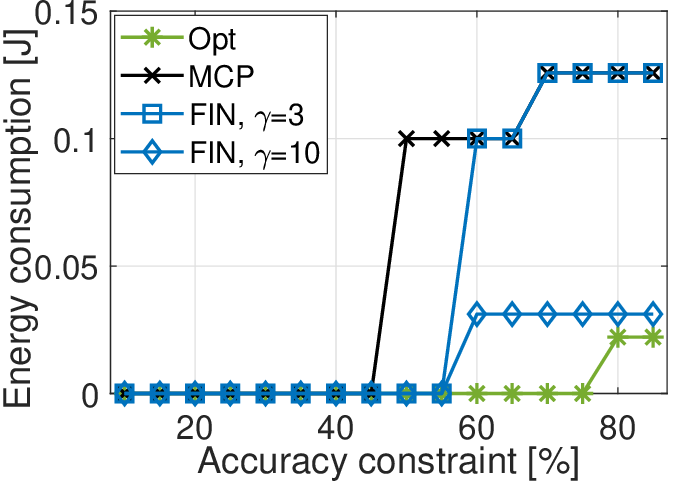}
}\\
    \vspace{-2mm}
        \subfigure[$\alpha^h{=}80$\%, compute energy\label{f:all_config_constraintd}]{
     \includegraphics[width=0.455\columnwidth]{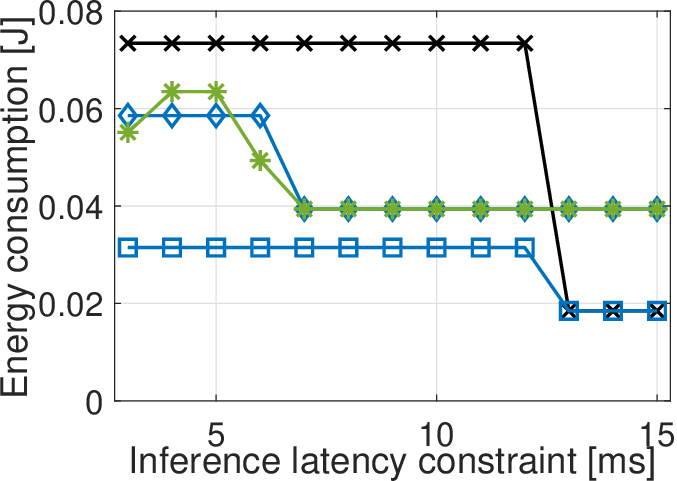}    
     }\hspace{-2mm}
     \subfigure[$\delta^h{=}$5\,ms, compute energy\label{f:all_config_constraintc}]{
         \includegraphics[width=0.455\columnwidth]{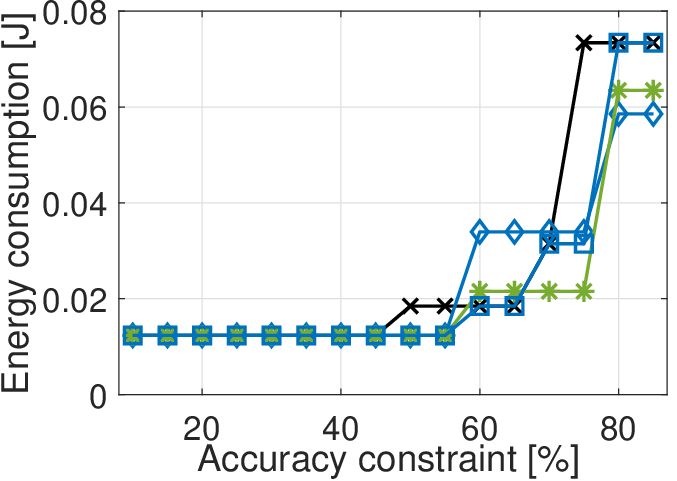} 
         }
    \vspace{-2mm}
    \caption{Computation and communication energy consumption of MCP, FIN, Opt for B-AlexNet   configurations, as the constraints vary.
    }
    \label{f:all_config_constraint}
\end{figure}

\noindent 
{\bf Performance of DNN deployments.}   
The total energy consumption of the B-AlexNet deployment configurations obtained using MCP, FIN, and Opt  is presented in Fig.\,\ref{f:tot_energy_balexnet}, as the inference accuracy and  latency constraints vary. The corresponding breakdown into communication and computation energy consumption  is depicted instead in Fig.\,\ref{f:all_config_constraint}. Fig.\,\ref{f:tot_energy_balexnet} shows that the energy consumption of the DNN configurations yielded by FIN ($\gamma{=}10$) is very close to the optimum and much less than MCP, while meeting the accuracy target $\alpha^h{=}80$\% and the latency target $\delta^h{=}5$\,ms. Also, for an extremely low value of $\gamma$ (namely, 3), FIN still outperforms MCP.   Our experiments have also revealed  that such performance ($\alpha^h{=}80$\%, $\delta^h{=}5$\,ms) is achieved by 
MCP, FIN ($\gamma{=}10$), and Opt  deploying (resp.) a set of [3,1,1], [2,1,2], and [1,2,2] blocks on [mobile, edge, cloud] nodes, and each employs exit-3 to meet the target accuracy. In this case, the deployment of the last two DNN blocks on the cloud reduces the energy consumption in FIN and Opt configuration as compared to MCP. For a less stringent latency requirement ($\alpha^h{=}80$\%, $\delta^h{=}12$\,ms), MCP  deploys [1,4,0] while FIN ($\gamma{=}10$) and Opt both deploy  [5,0,0] blocks, as the larger target latency allows  keeping  all blocks on the mobile, which reduces  energy consumption. In summary, meeting a smaller inference latency target requires a split deployment that increases energy expenditure.

\begin{figure}
    \centering
%
      \subfigure[$\alpha^h{=}85$\,\%\label{f:tot_energy_lenetb}]{
         \includegraphics[width=0.45\columnwidth]{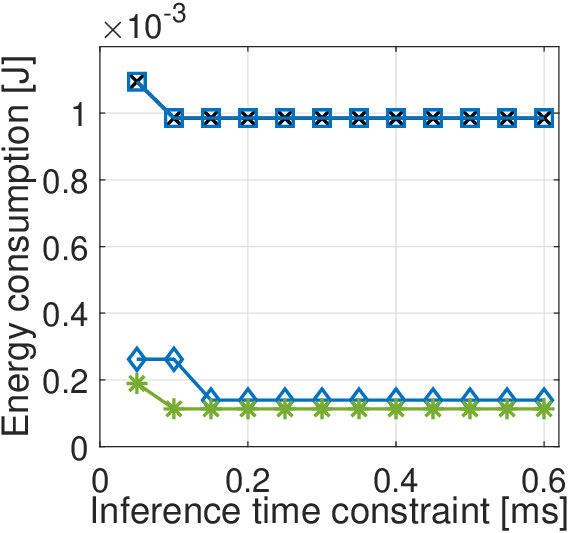}  
        }\hspace{-2mm}
     \subfigure[$\delta^h{=}$0.1\,ms\label{f:tot_energy_leneta}]{
 \includegraphics[width=0.45\columnwidth]{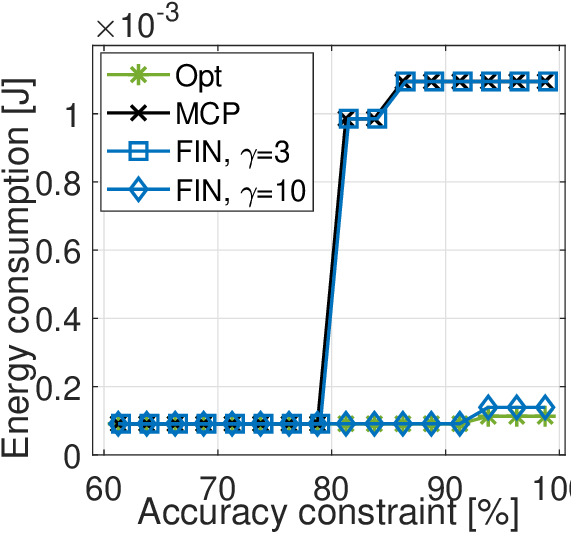}
 }
             \vspace{-2mm}
    \caption{Total energy consumption of the 
    B-LeNet  configurations obtained through Opt, MCP, and FIN  ($\gamma{=}3,10$),  as the target inference latency and accuracy vary. }
    \label{f:tot_energy_resnet_lenet}
    \vspace{-5mm}
\end{figure}

\begin{table}[!thb]
\vspace{-2mm}
\caption{Execution-time\,[ms] taken by MCP and FIN ($\gamma=3, 10$)  for finding the deployment configuration of each DNN\vspace{-3mm}}
\label{t:time_complexity}
\begin{center}
\small
\begin{tabular}{|c|c|c|c|}
    \hline
    {\bf Model}&{\bf MCP} &\multicolumn{2}{|c|}{\bf FIN}\\
    \cline{3-4}
    & &$\gamma{=}3$&$\gamma{=}10$ \\
    \hline
   B-AlexNet&0.591& 0.892 & 2.450\\ \hline
   B-ResNet&0.545& 0.657 & 1.158   \\ \hline
   B-LeNet&0.243  & 0.461  & 0.816\\ \hline
\end{tabular}
\end{center}
\vspace{-5mm}
\end{table}

\begin{figure*}[!htb]
\centering    
         \includegraphics[width=0.205\textwidth]{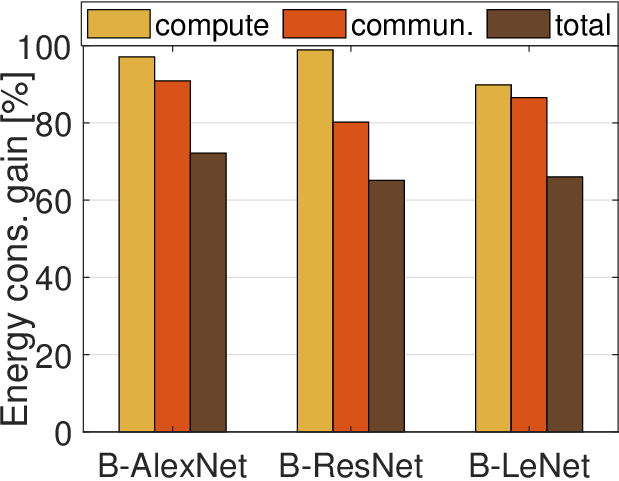}
         \hspace{0.5mm}
         \includegraphics[width=0.205\textwidth]{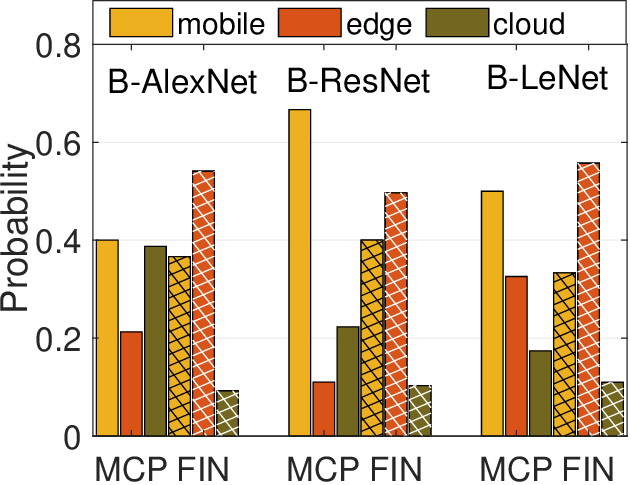}  
          \hspace{0.51mm}
         \includegraphics[width=0.125\textwidth]{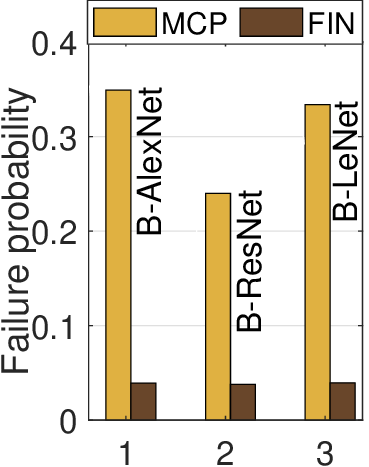}
          \hspace{0.5mm}
\includegraphics[width=0.405\textwidth]{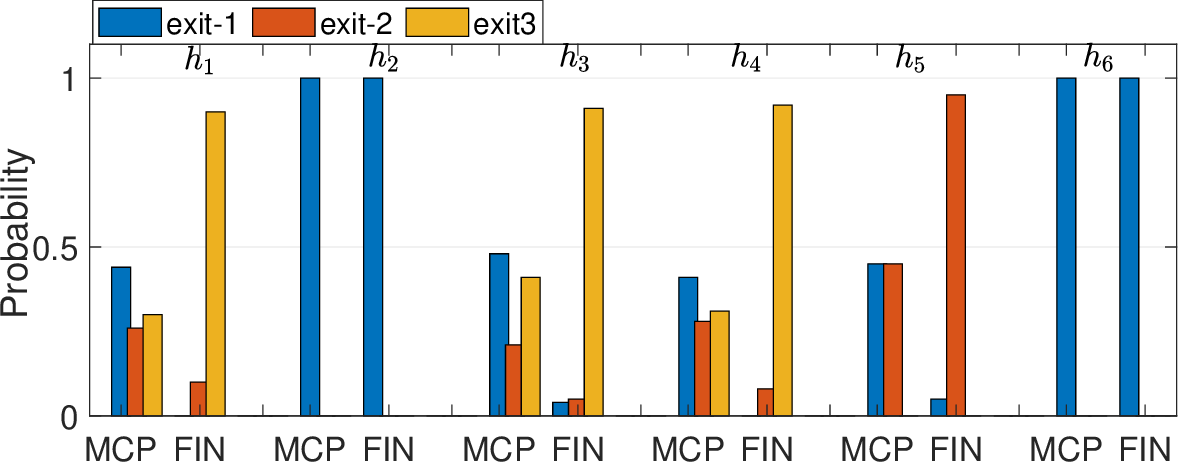}
    \vspace{-2mm}
    \caption{Performance in the multi-application scenario: Energy consumption gain through FIN  over MCP (left); probability of DNN block deployment on the multi-tier  nodes in MCP and FIN (center-left),  that the selected configuration fails to meet the constraints (center-right), and of DNN exit point for inference (right). For FIN, we set $\gamma{=}10$.}
\label{f:multi_model}
    \vspace{-5mm}
\end{figure*}

This is confirmed by Fig.\,\ref{f:all_config_constraint}(a)(c):    for a lower target inference latency,  communication energy consumption grows, as a split deployment is needed. Similarly, Fig.\,\ref{f:all_config_constraint}(b)(d)  underline that, with a higher value of accuracy constraint, the best configurations incur higher computation and communication energy: indeed, inspecting the resulting deployments, it emerges that they require later exit (exit-3) and split deployment. 
Comparing FIN to its benchmarks,  two main aspects are evident. First, even for a moderate value of~$\gamma{=}10$, FIN  virtually always matches the optimum and significantly outperforms MCP in all cases.  
When $\gamma$ drops to~$3$, the communication and computation energy expenditures diverge: the former (Fig.\,\ref{f:all_config_constraint}(a)(b)) deteriorates significantly; the latter (Fig.\,\ref{f:all_config_constraint}(c)(d)) remains remarkably low and stays close to the optimum. This suggests that communication energy is {\em harder} to minimize than computation energy, owing to the  complexity of the scenarios we target. Importantly, even when~$\gamma{=}3$ -- which, it is worth remarking, is extremely low -- FIN can match MCP. 
We have similarly evaluated the B-ResNet (omitted for brevity) and the B-LeNet  deployments (Fig.\,\ref{f:tot_energy_resnet_lenet}). 
Looking at FIN's  energy consumption,  a similar effect to Fig.\,\ref{f:tot_energy_balexnet} emerges, with FIN now  matching the optimum for sufficiently high $\gamma$ even more closely than in the case of B-AlexNet.

Table\,\ref{t:time_complexity} lists the overall execution times taken (on average) by MCP and FIN ($\gamma{=}3, 10$)  for obtaining the deployment configuration of the B-AlexNet, B-ResNet, and B-LeNet models, using a Lenovo ThinkPad P1 Gen 3 with i7-10750H CPU (2.6\,GHz, 32\,GB RAM).  The execution time of FIN is shorter than twice that of MCP for $\gamma{=}3$ and shorter than 5 times for $\gamma{=}10$.  Overall, the execution time of FIN is less than 2.5\,ms.

\noindent 
{\bf Multi-application scenario.}
We now apply FIN to the deployment of the  six applications listed in Table\,\ref{t:application}.  
Using the pre-trained DNN models for inference,  
we investigate the impact of an increasing number of users on the system performance.
We consider that 0.5\%  of the edge and cloud computing resources are available for each of the applications' inference execution. The application requirements, [inference latency [ms], accuracy [\%]], are set to [5,55], [5, 55], and [0.1, 93] for $h_{1-2}$, $h_{3-4}$, and $h_{5-6}$ applications, respectively. The energy consumption (computing and communication) gain provided by FIN ($\gamma{=}10$) and MCP, for the running applications $h_{1-6}$ that require one-inference-per-second-per-user, is shown in Fig.\,\ref{f:multi_model} (the size of this scenario renders obtaining the optimum impractical).  Notice how FIN  deploys the DNN model configuration that entails an overall energy consumption that is 65\%--70\% of the benchmark  for all the considered DNNs (Fig.\,\ref{f:multi_model}(left)).  
Also, the MCP approach leans towards more deployment on the mobile and cloud side, whereas FIN takes full advantage of all tiers and, in particular, of the edge (Fig.\,\ref{f:multi_model}(center-left)). Not only does FIN achieve better energy efficiency, but it also surpasses MCP in terms of success probability across all applications (Fig.\,\ref{f:multi_model}(center-right)). In contrast to MCP's high failure probability (over 30\% for B-Alexnet and B-Lenet and 20\% for B-Resnet), FIN sees less than 5\% of users failing to meet the latency and accuracy constraints. Consistently, 
Fig.\,\ref{f:multi_model}(right) shows that, with high probability and unlike MCP, FIN can deploy the applications blocks all the way to exit-3 whenever required, while it rightfully enables the earliest exit split whenever the accuracy constraint allows it
 (e.g., for $h_2$ and $h_6$).


\section{Related Work\label{sec:relwork}} 
Our work lies at the intersection of two major  fields, namely, model split (or, partitioning) and resource-aware ML.

\noindent
{\bf Early exit and model splitting~\cite{matsubara2022split}:} 
Early exit models have been introduced in~\cite{alexnet_ee}, which also raises the issue of how to place the early-exit layers, i.e., how to make the DNN topology {\em branchy}. In the same context, \cite{kang2017neurosurgeon}~tackles distributed scenarios and seeks to adapt the placement of exit layers to the available resources in the near-edge, edge, and cloud segments of the network. \cite{yang2022edge}~pursues a similar approach in IoT scenarios, minimizing the usage of edge resources. More recent work~\cite{lee2023wireless} performs DNN splitting in real time, with the aim of adapting to changes in channel conditions. The recent work~\cite{fan2023joint} widens the focus and accounts for the location of the users that need the inference task. 

Earlier approaches seek to split DNNs at naturally-occurring suitable locations, a.k.a. {\em bottlenecks}~\cite{bucilua2006model,chiang2021optimal}. If no bottleneck is available, the related problem of {\em bottleneck injection} arises. The goal is to 
change the topology of the DNN with the aim of creating suitable points to insert an early-exit layer. These techniques have been pioneered by~\cite{matsubara2019distilled,eshratifar2019bottlenet}, and often use pairs of encoder and decode layers. Bottleneck injection can be performed in a content-aware fashion, as in~\cite{lee2021splittable,matsubara2022bottlefit}. 
Also, \cite{zhang_wowmom23} underlines that collaborative DNN partitioning and task
offloading in resource-constrained edge-IoT network can meet the DNN inference deadline requirements.

\noindent 
{\bf Resource-aware ML} is, broadly speaking, concerned with adapting the distributed ML task to perform (whether it is training or inference) and the available resources. Works in this field  often focus on selecting the best nodes to exploit, accounting for their speed~\cite{wang2019adaptive}, size of local dataset~\cite{malandrino2021federated}, and feature-richness thereof~\cite{wu2021fast}, as well as any communication issues they may experience~\cite{zhou2021communication}. A more recent trend, closer in spirit to model partitioning, consists in  changing the learning task to fit the available resources, e.g., by selecting the most appropriate model~\cite{paissan2022scalable}.

A related trend is reliability in distributed ML. The main goal of this line of work is ensuring that all nodes involved in the ML task provide timely and high-quality updates, despite communication issues~\cite{ang2020robust} and the presence of malicious nodes~\cite{li2020learning}. Reliability might be at odd with fairness issues, and a balance between the two goals is sought in~\cite{li2021ditto}.

In summary, to the best of our knowledge, our work is the first to jointly tackle (i) how ML model splitting should be performed and (ii) where the different model blocks should be deployed, (iii)  for models with early exits as well as in the presence of inference requirements and constraints on the computational and networking resources in multi-tier  systems. 

\section{Conclusions\label{sec:conclusions}}
This paper addresses the problem of allocating sections of multi-branched dynamic DNNs to nodes in  mobile-edge-cloud systems. By means of a multi-stage graph-modeling approach, we solve the problem of minimizing the inference energy cost while matching the inference target requirements to the constrained nodes' resources. 
Our algorithmic solution, named FIN, to  this (NP-hard) problem leverages a further manipulation of the graph model to yield a low-complexity, yet effective, solution strategy.  
The results show that FIN closely matches the optimum and, by enabling effective split deployments and leveraging at best the nodes of the multi-tiered network,   reduces by over 65\% the inference energy consumption with respect to our benchmark. Future work will optimize the allocation of bandwidth and computational resources across different DNN-based applications in a  multi-tier system. 

\section*{Acknowledgments}
This work was supported by the European Commission through Grant No.\,101095890 (PREDICT-6G project),  Grant No.\,101096379 (CENTRIC project), and Grant No.\,101095363 (ADROIT6G project), and by the European Union under the Italian National Recovery and Resilience Plan (NRRP) of NextGenerationEU, partnership on “Telecommunications of the Future” (PE0000001 - program “RESTART”).

\bibliographystyle{IEEEtran}
\bibliography{refs.bib}

\begin{thebibliography}{10}
\providecommand{\url}[1]{#1}
\csname url@samestyle\endcsname
\providecommand{\newblock}{\relax}
\providecommand{\bibinfo}[2]{#2}
\providecommand{\BIBentrySTDinterwordspacing}{\spaceskip=0pt\relax}
\providecommand{\BIBentryALTinterwordstretchfactor}{4}
\providecommand{\BIBentryALTinterwordspacing}{\spaceskip=\fontdimen2\font plus
\BIBentryALTinterwordstretchfactor\fontdimen3\font minus
  \fontdimen4\font\relax}
\providecommand{\BIBforeignlanguage}[2]{{%
\expandafter\ifx\csname l@#1\endcsname\relax
\typeout{** WARNING: IEEEtran.bst: No hyphenation pattern has been}%
\typeout{** loaded for the language `#1'. Using the pattern for}%
\typeout{** the default language instead.}%
\else
\language=\csname l@#1\endcsname
\fi
#2}}
\providecommand{\BIBdecl}{\relax}
\BIBdecl

\bibitem{li2021survey}
Z.~Li, F.~Liu, W.~Yang, S.~Peng, and J.~Zhou, ``A survey of convolutional
  neural networks: analysis, applications, and prospects,'' \emph{IEEE
  transactions on neural networks and learning systems}, 2021.

\bibitem{nassif2019speech}
A.~B. Nassif, I.~Shahin, I.~Attili, M.~Azzeh, and K.~Shaalan, ``Speech
  recognition using deep neural networks: A systematic review,'' \emph{IEEE
  access}, vol.~7, pp. 19\,143--19\,165, 2019.

\bibitem{goldberg2022neural}
Y.~Goldberg, \emph{Neural network methods for natural language
  processing}.\hskip 1em plus 0.5em minus 0.4em\relax Springer Nature, 2022.

\bibitem{azimi2017hich}
I.~Azimi, A.~Anzanpour, A.~M. Rahmani, T.~Pahikkala, M.~Levorato, P.~Liljeberg,
  and N.~Dutt, ``Hich: Hierarchical fog-assisted computing architecture for
  healthcare iot,'' \emph{ACM Transactions on Embedded Computing Systems},
  2017.

\bibitem{zhang_wowmom23}
X.~Zhang, M.~Mounesan, and S.~Debroy, ``{EFFECT-DNN:} energy-efficient edge
  framework for real-time {DNN} inference,'' in \emph{Proc. {IEEE} {WoWMoM}},
  Boston, MA, USA, June 2023, pp. 10--20.

\bibitem{deng2020model}
L.~Deng, G.~Li, S.~Han, L.~Shi, and Y.~Xie, ``Model compression and hardware
  acceleration for neural networks: A comprehensive survey,'' \emph{Proceedings
  of the IEEE}, vol. 108, no.~4, pp. 485--532, 2020.

\bibitem{luo2021resource}
Q.~Luo, S.~Hu, C.~Li, G.~Li, and W.~Shi, ``Resource scheduling in edge
  computing: A survey,'' \emph{IEEE Communications Surveys \& Tutorials},
  vol.~23, no.~4, pp. 2131--2165, 2021.

\bibitem{matsubara2022split}
Y.~Matsubara, M.~Levorato, and F.~Restuccia, ``Split computing and early
  exiting for deep learning applications: Survey and research challenges,''
  \emph{ACM Computing Surveys}, vol.~55, no.~5, pp. 1--30, 2022.

\bibitem{kang2017neurosurgeon}
Y.~Kang, J.~Hauswald, C.~Gao, A.~Rovinski, T.~Mudge, J.~Mars, and L.~Tang,
  ``Neurosurgeon: Collaborative intelligence between the cloud and mobile
  edge,'' \emph{ACM SIGARCH Computer Architecture News}, 2017.

\bibitem{yang2022edge}
Y.-T. Yang and H.-Y. Wei, ``Edge--iot computing and networking resource
  allocation for decomposable deep learning inference,'' \emph{IEEE Internet of
  Things Journal}, vol.~10, no.~6, pp. 5178--5193, 2022.

\bibitem{miao2020adaptive}
W.~Miao, Z.~Zeng, L.~Wei, S.~Li, C.~Jiang, and Z.~Zhang, ``Adaptive dnn
  partition in edge computing environments,'' in \emph{IEEE ICPADS}, 2020.

\bibitem{lee2023wireless}
J.~Lee, H.~Lee, and W.~Choi, ``Wireless channel adaptive dnn split inference
  for resource-constrained edge devices,'' \emph{IEEE Communications Letters},
  2023.

\bibitem{fan2023joint}
W.~Fan, L.~Gao, Y.~Su, F.~Wu, and Y.~Liu, ``Joint dnn partition and resource
  allocation for task offloading in edge-cloud-assisted iot environments,''
  \emph{IEEE Internet of Things Journal}, 2023.

\bibitem{alexnet_ee}
S.~Teerapittayanon \emph{et~al.}, ``Branchynet: Fast inference via early
  exiting from deep neural networks,'' in \emph{IEEE ICPR}, 2016.

\bibitem{ee_sample_alexnet}
R.~Dong, Y.~Mao, and J.~Zhang, ``Resource-constrained edge ai with early exit
  prediction,'' \emph{Journal of Communications and Information Networks},
  vol.~7, no.~2, pp. 122--134, Jun. 2022.

\bibitem{laskaridis2021adaptive}
S.~Laskaridis, A.~Kouris, and N.~D. Lane, ``Adaptive inference through
  early-exit networks: Design, challenges and directions,'' in
  \emph{Proceedings of the 5th International Workshop on Embedded and Mobile
  Deep Learning}, 2021, pp. 1--6.

\bibitem{mcp1}
G.~Xue, A.~Sen, W.~Zhang, J.~Tang, and K.~Thulasiraman, ``Finding a path
  subject to many additive {QoS} constraints,'' \emph{{IEEE/ACM} Transactions
  on Networking}, vol.~15, no.~1, pp. 201--211, Feb. 2007.

\bibitem{GOUVEIA201922}
L.~Gouveia, M.~Leitner, and M.~Ruthmair, ``Layered graph approaches for
  combinatorial optimization problems,'' \emph{Computers \& Operations
  Research}, vol. 102, pp. 22--38, Feb. 2019.

\bibitem{GOUVEIA2017908}
------, ``Extended formulations and branch-and-cut algorithms for the
  black-and-white traveling salesman problem,'' \emph{European Journal of
  Operational Research}, vol. 262, no.~3, pp. 908--928, Nov. 2017.

\bibitem{Multilayer1}
S.~Yang, F.~Li, S.~Trajanovski, X.~Chen, Y.~Wang, and X.~Fu, ``Delay-aware
  virtual network function placement and routing in edge clouds,'' \emph{{IEEE}
  Transactions on Mobile Computing}, 2021.

\bibitem{hwang1992steiner}
F.~K. Hwang \emph{et~al.}, ``Steiner tree problems,'' \emph{Networks}, 1992.

\bibitem{xue2007finding}
G.~Xue, A.~Sen, W.~Zhang, J.~Tang, and K.~Thulasiraman, ``Finding a path
  subject to many additive qos constraints,'' \emph{IEEE/ACM Transactions on
  networking}, 2007.

\bibitem{AlexNet1}
\BIBentryALTinterwordspacing
O.~Elharrouss, Y.~Akbari, N.~Almaadeed, and S.~Al-Maadeed, ``Backbones-review:
  Feature extraction networks for deep learning and deep reinforcement learning
  approaches,'' Jun. 2022. [Online]. Available:
  \url{https://arxiv.org/abs/2206.08016}
\BIBentrySTDinterwordspacing

\bibitem{alexnet_vid_sum}
J.~Lei, Q.~Luan, X.~Song, X.~Liu, D.~Tao, and M.~Song, ``Action parsing-driven
  video summarization based on reinforcement learning,'' \emph{IEEE
  Transactions on Circuits and Systems for Video Technology}, 2019.

\bibitem{alexnet_activity1}
F.~Serpush and M.~Rezae, ``Complex human action recognition using a
  hierarchical feature reduction and deep learning-based method,'' \emph{{SN}
  Computer Science}, vol.~2, no.~94, pp. 1--15, Feb. 2021.

\bibitem{alexnet_activity2}
A.~Ullah, J.~Ahmad, K.~Muhammad, M.~Sajjad, and S.~W. Baik, ``Action
  recognition in video sequences using deep bi-directional {LSTM} with {CNN}
  features,'' \emph{{IEEE} Access}, vol.~6, pp. 1155--1166, 2018.

\bibitem{alexnet_activity3}
S.~Darafsh, S.~S. Ghidary, and M.~S. Zamani, ``Real-time activity recognition
  and intention recognition using a vision-based embedded system,''
  \emph{CoRR}.

\bibitem{resnet}
K.~He, X.~Zhang, S.~Ren, and J.~Sun, ``Deep residual learning for image
  recognition,'' 2015.

\bibitem{LeNet}
Y.~Lecun, L.~Bottou, Y.~Bengio, and P.~Haffner, ``Gradient-based learning
  applied to document recognition,'' \emph{Proceedings of the IEEE}, 1998.

\bibitem{mnist}
Y.~LeCun, C.~Cortes, and C.~Burges, ``Mnist handwritten digit database,''
  \emph{ATT Labs}, 2010.

\bibitem{emnist}
G.~Cohen, S.~Afshar, J.~Tapson, and A.~van Schaik, ``Emnist: an extension of
  mnist to handwritten letters,'' \emph{arXiv preprint arXiv:1702.05373}, 2017.

\bibitem{compute_par}
F.~Malandrino, C.~F. Chiasserini, and G.~di~Giacomo, ``Efficient distributed
  {DNN}s in the mobile-edge-cloud continuum,'' \emph{{IEEE/ACM} Transactions on
  Networking (early access)}, pp. 1--15, Nov. 2022.

\bibitem{commun_par1}
F.~Jalali, K.~Hinton, R.~Ayre, T.~Alpcan, and R.~S. Tucker, ``Fog computing may
  help to save energy in cloud computing,'' \emph{{IEEE} Journal on Selected
  Areas in Communications}, 2016.

\bibitem{commun_par2}
Y.~Li, A.-C. Orgerie, I.~Rodero, B.~L. Amersho, M.~Parashar, and J.-M. Menaud,
  ``End-to-end energy models for edge cloud-based {IoT} platforms: Application
  to data stream analysis in {IoT},'' \emph{Future Generation Computer
  Systems}, vol.~87, pp. 667--678, Oct. 2018.

\bibitem{commun_par3}
L.~Sun, H.~Deng, R.~K. Sheshadri, W.~Zheng, and D.~Koutsonikolas,
  ``Experimental evaluation of {WiFi} active power/energy consumption models
  for smartphones,'' \emph{{IEEE} Transactions on Mobile Computing}, vol.~16,
  no.~1, pp. 115--129, Mar. 2017.

\bibitem{bucilua2006model}
C.~Buciluǎ, R.~Caruana, and A.~Niculescu-Mizil, ``Model compression,'' in
  \emph{ACM SIGKDD}, 2006.

\bibitem{chiang2021optimal}
C.-H. Chiang, P.~Liu, D.-W. Wang, D.-Y. Hong, and J.-J. Wu, ``Optimal branch
  location for cost-effective inference on branchynet,'' in \emph{IEEE Big
  Data}, 2021.

\bibitem{matsubara2019distilled}
Y.~Matsubara, S.~Baidya, D.~Callegaro, M.~Levorato, and S.~Singh, ``Distilled
  split deep neural networks for edge-assisted real-time systems,'' in
  \emph{ACM HotEdgeVideo}, 2019.

\bibitem{eshratifar2019bottlenet}
A.~E. Eshratifar, A.~Esmaili, and M.~Pedram, ``Bottlenet: A deep learning
  architecture for intelligent mobile cloud computing services,'' in
  \emph{IEEE/ACM ISLPED}, 2019.

\bibitem{lee2021splittable}
J.~C. Lee, Y.~Kim, S.~Moon, and J.~H. Ko, ``A splittable dnn-based object
  detector for edge-cloud collaborative real-time video inference,'' in
  \emph{IEEE AVSS}, 2021.

\bibitem{matsubara2022bottlefit}
Y.~Matsubara, D.~Callegaro, S.~Singh, M.~Levorato, and F.~Restuccia,
  ``Bottlefit: Learning compressed representations in deep neural networks for
  effective and efficient split computing,'' in \emph{IEEE WoWMoM}, 2022.

\bibitem{wang2019adaptive}
S.~Wang, T.~Tuor, T.~Salonidis, K.~K. Leung, C.~Makaya, T.~He, and K.~Chan,
  ``Adaptive federated learning in resource constrained edge computing
  systems,'' \emph{IEEE Journal on Selected Areas in Communications}, 2019.

\bibitem{malandrino2021federated}
F.~Malandrino and C.~F. Chiasserini, ``{Federated learning at the network edge:
  When not all nodes are created equal},'' \emph{IEEE Communications Magazine},
  2021.

\bibitem{wu2021fast}
H.~Wu and P.~Wang, ``Fast-convergent federated learning with adaptive
  weighting,'' \emph{IEEE Transactions on Cognitive Communications and
  Networking}, 2021.

\bibitem{zhou2021communication}
Y.~Zhou, Q.~Ye, and J.~C. Lv, ``{Communication-Efficient Federated Learning
  with Compensated Overlap-FedAvg},'' \emph{IEEE Transactions on Parallel and
  Distributed Systems}, 2021.

\bibitem{paissan2022scalable}
F.~Paissan, A.~Ancilotto, A.~Brutti, and E.~Farella, ``Scalable neural
  architectures for end-to-end environmental sound classification,'' in
  \emph{IEEE ICASSP}, 2022.

\bibitem{ang2020robust}
F.~Ang, L.~Chen, N.~Zhao, Y.~Chen, W.~Wang, and F.~R. Yu, ``Robust federated
  learning with noisy communication,'' \emph{IEEE Transactions on
  Communications}, 2020.

\bibitem{li2020learning}
S.~Li, Y.~Cheng, W.~Wang, Y.~Liu, and T.~Chen, ``Learning to detect malicious
  clients for robust federated learning,'' \emph{arXiv preprint
  arXiv:2002.00211}, 2020.

\bibitem{li2021ditto}
T.~Li, S.~Hu, A.~Beirami, and V.~Smith, ``Ditto: Fair and robust federated
  learning through personalization,'' in \emph{ICML}, 2021.

\end{thebibliography}

\end{document}